\documentclass[pdflatex,sn-mathphys-num]{sn-jnl}

\usepackage{graphicx}%
\usepackage{multirow}%
\usepackage{amsmath,amssymb,amsfonts}%
\usepackage{amsthm}%
\usepackage{mathrsfs}%
\usepackage[title]{appendix}%
\usepackage{xcolor}%
\usepackage{textcomp}%
\usepackage{manyfoot}%
\usepackage{booktabs}%
\usepackage{algorithm}%
\usepackage{algorithmicx}%
\usepackage{algpseudocode}%
\usepackage{listings}%

\usepackage{bm}
\usepackage{mathtools} 
\usepackage{physics}

\theoremstyle{thmstyleone}%
\theoremstyle{thmstyletwo}%

\theoremstyle{thmstylethree}%

\begin{document}

\title[Interatomic spin-orbit interaction in a $p$-orbital helical atomic chain]{Interatomic spin-orbit interaction in a $p$-orbital helical atomic chain}


\author[1]{\fnm{Takemitsu} \sur{Kato}}\email{tkato.take@gmail.com}

\author*[1,2]{\fnm{Yasuhiro} \sur{Utsumi}}\email{utsumi@phen.mie-u.ac.jp}

\author[3]{\fnm{Ora} \sur{Entin-Wohlman}}\email{orawohlman@gmail.com}
\author[3]{\fnm{Amnon} \sur{Aharony}}\email{aaharonyaa@gmail.com}
\equalcont{These authors contributed equally to this work.}

\affil*[1]{\orgdiv{Department of Physics Engineering, Faculty of Engineering}, \orgname{Mie University}, \orgaddress{\street{1577, Kurimamachiya-cho}, \city{Tsu}, \postcode{514-8507}, \state{Mie}, \country{Japan}}}

\affil[2]{\orgdiv{Department of Electrical and Electronic Engineering, Faculty of Engineering}, \orgname{Mie University}, \orgaddress{\street{1577, Kurimamachiya-cho}, \city{Tsu}, \postcode{514-8507}, \state{Mie}, \country{Japan}}}
\affil[3]{\orgdiv{School of Physics and Astronomy}, \orgname{Tel Aviv University}, \orgaddress{\city{Tel Aviv}, \postcode{6997801}, \country{Israel}}}


\abstract{
We derive the interatomic spin-orbit interaction (SOI) from a helical atomic chain composed of $p$-orbitals with intra-atomic SOI, which exhibits a helical state—a potential origin of the chiral-induced spin selectivity (CISS) effect. In this model, a strong crystal field in the tangential direction of the helix leads to the formation of energetically separated $\sigma$- and $\pi$-bands. In the second-order process, a spin in the $\sigma$-orbital virtually hops to the $\pi$-orbital, flips its direction due to intra-atomic SOI, and then hops back to the $\sigma$-orbital in the neighboring atom due to the misalignment of $p$-orbitals along the helix. This process induces an interatomic SOI in the $\sigma$-band, which takes the form of a Rashba-type SOI generated by an electric field normal to the helical axis. The magnitude of the SOI is proportional to the curvature, the hopping energy, the intra-atomic SOI energy, and inversely proportional to the crystal field strength. The second-order process also induces long-range second-nearest-neighbor hoppings. We analytically derive the spin-split band structure in the zero-torsion limit.
}




\maketitle

\section{Introduction}

The chirality-induced spin selectivity (CISS) effect is a spin dependent phenomenon specific to chiral materials~\cite{Bloom2024}. 
The prototypical setup for CISS involves electron transfer through a single helical molecule, such as a DNA molecule or a helicene molecule (see, e.g.,\cite{Mishra2020,Singh2025}): when a spin is injected into a molecule, it is selectively transmitted depending on the chirality of molecule. 
Although numerous theoretical proposals have been made for this setup, achieving quantitative agreement with experiments remains an open challenge~\cite{EversAM2022}.
One possible reason for this discrepancy lies in the complexity of the molecules, which has led to continuous efforts to develop simplified toy models that capture the essential physics of this spin-dependent phenomenon and potentially explain experimental results (see, e.g.,~\cite{Korytar2024,diventra2025,goebel2025}). 

In this context, most studies assume that the CISS effect originates from either inter-atomic spin-orbit interaction (SOI) or intra-atomic SOI.
A $p$-orbital helical atomic chain with intra-atomic SOI~\cite{KatoJCP2023,UtsumiIJC2022,UtsumiPRB2020} serves as a minimal model for a two-terminal, two-orbital spin filter.
This model is simple enough to allow for analytical calculations on the band structure~\cite{UtsumiPRB2020,UtsumiIJC2022} and electron and spin states in a finite chain~\cite{UtsumiIJC2022}.

In this model, a strong crystal field along the helix energetically separates the $\sigma$- and $\pi$-bands.
In the $\pi$-bands, two helical states emerge within an energy window proportional to the curvature of the helix and the intra-atomic SOI energy~\cite{UtsumiPRB2020,UtsumiIJC2022,KatoJCP2023}.
In this paper, we demonstrate that in the $\sigma$-band, a Rashba-like inter-atomic SOI exists, whose strength depends on the curvature of the helix. 
This curvature-dependent interatomic SOI has been derived for carbon nanotubes and curved graphenes~\cite{HuertasHernandoPRB2006}, as well as in a CISS tight-binding model~\cite{VarelaPRB2016}.

The structure of this paper is as follows:
In Sec.~\ref{sec1}, we introduce the model Hamiltonian and a local coordinate system
that diagonalizes the crystal field along the tangential direction of the helix. 
In Sec.~\ref{sec2}, we perform the Schrieffer-Wolff transformation and derive the inter-atomic SOI. 
Then, after presenting numerical results in Sec.~\ref{sec3}, we summarize our findings in Sec.~\ref{sec4}.

\section{Model Hamiltonian: $p$-orbital helical atomic chain}
\label{sec1}

\begin{figure}
\begin{center}
\includegraphics[width=0.7 \columnwidth]{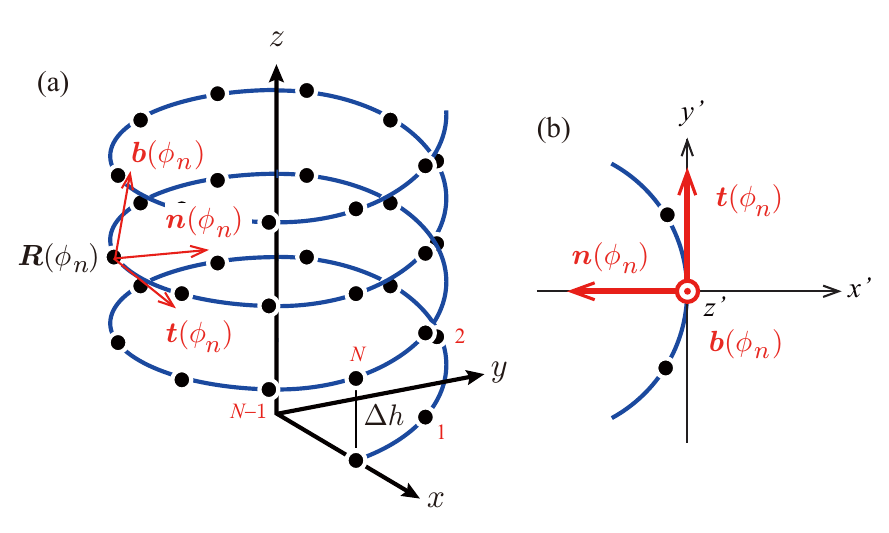}
\caption{
(a) Schematic picture of a helical atomic chain.
The model is a right-handed system in which the $z$-axis coincides with the helical axis. 
(b) Schematic picture of the local orthogonal coordinate system at site $n$. 
The basis set is chosen as $\{ -{\bm n}(\phi_n),{\bm t}(\phi_n),{\bm b}(\phi_n)\}$. 
}
\label{fig:dna}
\end{center}
\end{figure}

Figure~\ref{fig:dna} (a) shows the schematic picture of our model. 
The position of an atom on the helical atomic chain is,
\begin{align} 
{\bm R}(\phi_n) = \left( R\cos(\phi_n), R\sin(p\phi_n), \Delta h  \phi_n/(2\pi) \right) . 
\end{align} 
where $R$, $n$, $\Delta h$ and $\phi_n = \Delta \phi \, n$ represent the radius, site number, the pitch, and the rotation angle around the z-axis, respectively. 
Here $p = +1 (-1)$ indicates the right (left)-handed helix. 
The angle between the neighboring atoms is $\Delta \phi = 2\pi/N$, where $N$ is the number of atoms in a single turn, i.e. a unit cell. 
The helix is conveniently treated in the Frenet-Serret frame, in which the tangent ${\bm t}$ (along the helix), normal ${\bm n}$, and bi-normal ${\bm b}$ vectors at a point on the helix are, 
\begin{align} 
{\bm t}(\phi) &= (- \kappa \sin (\phi), p \kappa \cos (\phi), |\tau| ) \ , 
\\ 
{\bm n}(\phi) &= (- \cos (\phi), - p \sin (\phi), 0 ) \ , 
\\ 
{\bm b}(\phi) &= {\bm t}(\phi) \times {\bm n}(\phi) = ( p |\tau| \sin (\phi), - |\tau| \cos (\phi), p \kappa ) \ , 
\label{tnb} 
\end{align} 
where the `normalized' curvature and torsion, $\kappa$ and $\tau$, are, 
\begin{align} 
\kappa &= \frac{R}{\sqrt{R^2 + [\Delta h/(2 \pi)]^2}} \eqqcolon \cos (\theta )\ , 
\;\; 
\tau = \frac{p \Delta h/(2 \pi)}{\sqrt{R^2 + [\Delta h/(2 \pi)]^2}} \eqqcolon p \sin (\theta) \ . 
\label{eqn:tau} 
\end{align} 

The Hamiltonian of the helical atomic chain with $M$ turns (unit cells) is given by
\begin{align}  
{\mathcal H}_{\rm mol} =& \sum_{n=1}^{MN} \biggl( -\tilde c^\dagger_{n+1}\bm J\otimes\sigma^{}_0\ \tilde c^{}_{n} +\mathrm{H.c.}   
+\Delta_{\mathrm{so}} \tilde c^\dagger_{n}\ \bm L \cdot \bm \sigma\ \tilde c^{}_{n}  
\notag\\ &  
+K_{\bm t} \tilde c^\dagger_{n} \left[\left(\bm t (\phi_n)\cdot\bm L\right)^2-I_3\right]\otimes\sigma^{}_0\ \tilde c^{}_{n} \biggr) .  
\label{fullH}  
\end{align}  
Here, the $n$-th atom hosts $p$ orbitals, and the vector of creation operators is  
\begin{equation}  
\tilde c^\dag_{n}=  
\left( \tilde c^\dag_{n;x\uparrow} \;  
\tilde c^\dag_{n;x\downarrow} \;  
\tilde c^\dag_{n;y\uparrow} \;  
\tilde c^\dag_{n;y\downarrow} \;  
\tilde c^\dag_{n;z\uparrow} \;  
\tilde c^\dag_{n;z\downarrow} \right),  
\end{equation}  
where $\tilde{c}^\dag_{n;o \sigma_s}$ ($o=p_x,p_y,p_z$) creates a spin-$\sigma_s$ electron in orbital $o$ at site $n$.  
The first term represents the electron hopping between nearest-neighbor atoms.  
The second term describes the intra-atomic SOI.  
The third term accounts for the crystal field along the tangential direction of the helix, which characterizes the helical structure of the molecule. 
The Hamiltonian is symmetric under time reversal, i.e.,  
\begin{equation}  
{\mathcal H}_{\rm mol} = \Theta {\mathcal H}_{\rm mol} \Theta^{-1} \, ,  
\end{equation}  
where the time-reversal operator is  
$\Theta = -i \sigma_y K$,  
and $K$ is the complex conjugation operator.

In the second term, $\Delta_{\mathrm{so}}$ is the SOI strength.  
The boldface vectors ${\bm \sigma}=(\sigma_x,\sigma_y,\sigma_z)$ and ${\bm L}=(L_x,L_y,L_z)$  
represent the vector of $2 \times 2$ Pauli matrices and the vector of orbital angular momentum operators, respectively.  
We write the basis functions of $p$-orbitals as $\ket{x}$, $\ket{y}$, and $\ket{z}$.  
The orbital angular momentum operators are written, for example, as $L_x=-i ( \ketbra{y}{z} - \ketbra{z}{y} )$.  
In matrix form, they are given by  
\begin{align}  
L_x=\begin{pmatrix}  
0&0&0\\0&0&-i\\0&i&0  
\end{pmatrix} , \;\;  
L_y=\begin{pmatrix}  
0&0&i\\0&0&0\\-i&0&0  
\end{pmatrix} , \;\;  
L_z=\begin{pmatrix}  
0&-i&0\\i&0&0\\0&0&0  
\end{pmatrix} .  
\end{align}  

This system satisfies the helical symmetry~\cite{OtsutoPRB2021,UtsumiIJC2022}.  
The matrix ${\bm J}$ in the first term of the Hamiltonian is a $3 \times 3$ matrix in orbital space, satisfying the following commutation relation:  
\begin{align}  
\left[ e^{i L_z p \Delta \phi} , {\bm J} \right]=0  \, , \label{helical_symmetry}  
\end{align}  
By also accounting for time-reversal symmetry, i.e.,  
$K {\bm J} K^{-1}={\bm J}$, the matrix ${\bm J}$ can be parameterized by three real numbers, $J$, $\alpha$, and $\varphi$, as  
\begin{align}  
{\bm J} =J \begin{pmatrix} \alpha \cos (p\varphi) &  -\alpha \sin (p\varphi)&0\\\alpha \sin (p\varphi)& \alpha \cos (p\varphi) &0\\0&0&1 \end{pmatrix} \, . \label{mJ}  
\end{align}  
In the following, we take the hopping energy as $J>0$.  
The matrix $\sigma^{}_0$ in the first term is a $2 \times 2$ identity matrix in spin space.  

In the second line of (\ref{fullH}), $K_{\bm t}$ represents the crystal field along the tangential direction.  
The symbol $I_3$ denotes the $3 \times 3$ identity matrix.  

Since the crystal field energy sets the largest energy scale in this model,
we first diagonalize it using a local orthogonal transformation:  
\begin{align}  
c^{}_n= O^{}_n \tilde{c}^{}_n \ ,  
\;\;\;  
O^{}_n=e^{i L^{}_x \theta^{}_{p}} e^{i L^{}_z p \, \phi^{}_{n}} \ ,   
\;\;\;   
\theta^{}_{p}= \frac{\pi}{2} (1-p) + p \, \theta \ . \label{On}  
\end{align}  
The orthonormal basis vectors of the local coordinate system are chosen as  
$ \{  -{\bm n}(\phi),{\bm t}(\phi),{\bm b}(\phi) \}$, see Fig.~\ref{fig:dna} (b).  
In this basis, the Hamiltonian is, 
\begin{align}  
{\mathcal H}_{\rm mol} =& \sum_{n=1}^{MN} \biggl( - c^\dagger_{n+1}  O^{}_{n+1} \bm J O^{\dagger}_{n} \otimes\sigma^{}_0\  c^{}_{n} +\mathrm{H.c.}   
+\Delta_{\mathrm{so}} c^\dagger_{n}\ O^{}_{n} \bm L O^{\dagger}_{n} \cdot \bm \sigma\ c^{}_{n}  
\notag\\ & +K_{\bm t} c^\dagger_{n} \left[\left(\bm t (\phi_n)\cdot O^{}_{n} \bm L O^{\dagger}_{n} \right)^2 - I_3\right] \otimes \sigma^{}_0\ c^{}_{n} \biggr), \label{fullH_}  
\end{align}  
where the crystal field is diagonalized, since the squared inner products, 
\begin{align}  
\left( -{\bm n}(\phi^{}_n)  \cdot O_n^{} {\bm L} O_n^{\dagger} \right)^2 = &  L^{2}_x = I_3 - \ketbra{x}{x} \, , \\  
\left( {\bm t}(\phi^{}_n) \cdot O_n^{} {\bm L} O_n^{\dagger} \right)^2 =& L^{2}_y = I_3 - \ketbra{y}{y} \, , \\  
\left( {\bm b}(\phi^{}_n)  \cdot O_n^{} {\bm L} O_n^{\dagger} \right)^2 =&  L^{2}_z = I_3 - \ketbra{z}{z} ,  
\end{align}  
are diagonal.

The orthogonal transformation (\ref{On}) induces the mixing of different $p$-orbitals at adjacent sites,  
leading to the modified hopping matrix  
$\tilde{\bm J}=O_{n+1} \bm J O_n^{\dagger}$,  
\begin{align}
\tilde{\bm J} =& J
e^{i L^{}_x \theta^{}_{p}}
\begin{pmatrix}
J_+ & -p J_- & 0 \\
p J_- & J_+  & 0 \\
0&0&J
\end{pmatrix}
e^{-i L^{}_x \theta^{}_{p}}
=
\tilde{\bm J}_\pi + \tilde{\bm J}_{\sigma} + \tilde{\bm J}_{\pi-\sigma}
\, , 
\end{align}
where  
$J_+ = J \alpha \cos (\delta \phi)$,  
$J_- = J \alpha \sin (\delta \phi)$,  
and  
$\delta \phi = \varphi - \Delta \phi$. 
In the right-hand side, 
\begin{align}
\tilde{\bm J}_\pi =& J_+ \ketbra{x}{x} - i J_- \tau  L_y + \left( J_+ \tau^2 + J \kappa^2 \right) \ketbra{z}{z} \, , 
\\
\tilde{\bm J}_{\sigma} =& \left( J_+ \kappa^2 + J \tau^2 \right) \ketbra{y}{y} \, ,
\\
\tilde{\bm J}_{\pi-\sigma} =& - i \mathcal J_- L_z + \mathcal J_+ P_x \, , 
\end{align}
where 
$\mathcal J_- = J_- \kappa$,  
and $\mathcal J_+ = (J - J_+) p \tau \kappa$. 
Here we introduced the operator of switching two orbitals, $P_x = \ketbra{z}{y}+\ketbra{y}{z}$,  

The rotation angle $\Delta \phi$ arises due to the misalignment of the local orthogonal coordinates at adjacent sites.  
When this rotational effect is compensated by the parameter $\varphi$, i.e.,  
$\delta \phi = 0$,  
the hopping matrix becomes diagonal, $O_{n+1} \bm J O_n^{-1} = J {\rm diag} (\alpha, \alpha, 1)$.  
Furthermore, when $\alpha = 1$, the ideal condition is realized, leading to the formation of the helical state~\cite{UtsumiPRB2020,UtsumiIJC2022,KatoJCP2023}.

The intra-atomic SOI term is,  
\begin{align}  
O^{}_{n} \bm L O^{-1}_{n} \cdot \bm \sigma\ = - L_x \left(  {\bm n}(\phi_n)  \cdot{\bm \sigma} \right) + L_y \left( {\bm t}(\phi_n) \cdot {\bm \sigma} \right) + L_z \left( {\bm b}(\phi_n) \cdot {\bm \sigma} \right) \eqqcolon \bm L \cdot \tilde{\bm \sigma} (\phi_n) \, ,  
\end{align}  
where the Pauli matrices are rotated (Appendix~\ref{secA}) as,  
\begin{align}  
- {\bm n}(\phi_n)  \cdot{\bm \sigma} =&   
\begin{pmatrix}  
0 & e^{-i p \phi_n } \\  
e^{i p \phi_n } & 0 \\  
\end{pmatrix} \eqqcolon \tilde{\sigma}_x(\phi_n) \label{eqn:tsx}  
\, ,  
\\  
{\bm t}(\phi_n)  \cdot{\bm \sigma} =&   
p  
\begin{pmatrix}  
\tau & -i e^{-i p \phi_n } \kappa \\  
i e^{i p \phi_n } \kappa  & -\tau \\  
\end{pmatrix} \eqqcolon \tilde{\sigma}_y(\phi_n) \label{eqn:tsy}  
\, ,  
\\  
{\bm b}(\phi_n)  \cdot{\bm \sigma} =&   
p  
\begin{pmatrix}  
\kappa & i e^{-i p \phi_n } \tau \\  
-e^{i p \phi_n } \tau & -\kappa \\  
\end{pmatrix} \eqqcolon \tilde{\sigma}_z(\phi_n) \label{eqn:tsz}  
\, .    
\end{align}  
Summarizing the above, the Hamiltonian in the local coordinate system reads,  
\begin{align}  
{\mathcal H}_{\rm mol}=\sum_{n=1}^{MN} \left( -c_{n+1}^\dag\tilde{\bm J}\otimes\sigma _0\ c_n +\mathrm{H.c.}+\Delta_{\mathrm{so}} c_{n}^\dag {\bm L}\cdot \tilde{\bm\sigma}(\phi_n) c_n-K_{\bm t} c_{n}^\dag \ketbra{y}{y} \otimes\sigma _0\ c_n \right) \, .  
\end{align}  

\section{Inter-atomic SOI}
\label{sec2}

The $p_y$ orbitals in the local coordinate system form a $\sigma$-band, while the $p_x$ and $p_z$ orbitals form a $\pi$-band, which are energetically separated by $K_{\bm t}$. 
We partition the Hamiltonian as follows: 
\begin{align}
{\mathcal H}_{\rm mol}  &= {\mathcal H}_\pi + {\mathcal H}_\sigma + {\mathcal H}_{\pi-\sigma} \ , 
\end{align}
where
\begin{align}
{\mathcal H}_{\pi}  &= 
\sum_{n=1}^{MN} \left( -c_{n+1}^\dag \tilde{\bm J}_\pi \otimes \sigma _0\ c_n + \mathrm{H.c.} + \Delta_{\mathrm{so}} c_{n}^\dag {L}_y \tilde{\sigma}_y c_n \right) \, ,
\\
{\mathcal H}_{\sigma}  &= 
\sum_{n=1}^{MN} \left( -c_{n+1}^\dag\tilde{\bm J}_\sigma \otimes \sigma _0\ c_n + \mathrm{H.c.} - K_{\bm t} \, c_{n}^\dag \ketbra{y}{y} \otimes \sigma _0\ c_n \right) \, ,
\\
{\mathcal H}_{\pi - \sigma}  &= 
\sum_{n=1}^{MN} \left( -c_{n+1}^\dag\tilde{\bm J}_{\pi-\sigma} \otimes \sigma _0\ c_n + \mathrm{H.c.} + \Delta_{\mathrm{so}} c_{n}^\dag \left( {L}_x \tilde{\sigma}_x + {L}_z \tilde{\sigma}_z \right) c_n \right) \, ,
\end{align}

In the $\sigma$-band, the inter-atomic SOI arises due to the mixing between the $\sigma$ and $\pi$ orbitals, which we will derive based on the Schrieffer-Wolff transformation~\cite{Schrieffer1966}, the second-order (quasi-)degenerate perturbation in ${\mathcal H}_{\pi-\sigma}$. 
Let $\ket{\sigma_i}$ and $\ket{\sigma_j}$ denote the single-particle eigenstates of the Hamiltonian ${\mathcal H}_{\sigma}$ satisfying 
${\mathcal H}_{\sigma} \ket{\sigma_{i(j)}} = \epsilon_{i(j)} \ket{\sigma_{i(j)}}$. 
For $K_{\bm t} \gg J , \Delta_{\rm so}$, 
\begin{align}
\frac{1}{2} \mel{\sigma_i}{ {\mathcal H}_{\pi-\sigma} 
\left( \frac{1}{\epsilon_i - {\mathcal H}_{\pi}} + \frac{1}{\epsilon_j - {\mathcal H}_{\pi}} \right)
{\mathcal H}_{\pi-\sigma} }{\sigma_j}
\approx
\mel{\sigma_i}{ - \frac{ {\mathcal H}_{\pi-\sigma}^2 }{K_{\bm t}} }{\sigma_j}
\, ,
\end{align}
and thus the effective Hamiltonian becomes, 
${\mathcal H}_{\sigma}^\prime = {\mathcal H}_{\sigma} - {\mathcal H}_{\pi-\sigma}^2/K_{\bm t}$. 
The non-vanishing components of ${\mathcal H}_{\pi-\sigma}^2$ are,  
\begin{align}
{\mathcal H}_{\pi - \sigma}^2 =& 
\sum_{n=1}^{MN}
c_{n}^\dag \left( \left[ \tilde{\bm J}_{\pi-\sigma} , \tilde{\bm J}_{\pi-\sigma}^\dagger \right]_+ \otimes \sigma _0 + \Delta_{\mathrm{so}}^2 \left({L}_x \tilde{\sigma}_x + {L}_z \tilde{\sigma}_z \right)^2 \right) c_n 
\notag \\ &+
c_{n+1}^\dag \left[ \tilde{\bm J}_{\pi-\sigma} , \Delta_{\mathrm{so}} \left({L}_x \tilde{\sigma}_x + {L}_z \tilde{\sigma}_z \right) \right]_+ c_n 
+\mathrm{H.c.}
\notag \\ &+
c_{n+2}^\dag \ \tilde{\bm J}_{\pi-\sigma} \tilde{\bm J}_{\pi-\sigma} \ c_n 
+\mathrm{H.c.} \, ,
\label{eqn:Heff_intermediate}
\end{align}
where $\left[ A,B \right]_+ = AB + BA$ represents the anti-commutator. 
Then the straightforward calculations lead to (Appendix~\ref{secB}), 
\begin{align}
{\mathcal H}_{\sigma}^{\prime} = &
\sum_{n=1}^{MN} 
c_{n+1;y}^\dagger \left[ - \left( J_+ \kappa^2 + J \tau^2 \right) {\sigma}_{0} + \frac{2 i \mathcal J_- \Delta_{\rm so}}{K_{\bm t}} \tilde{\sigma}_{z}(\phi_n) \right]
c_{n;y}+ 
\mathrm{H.c.}
\notag \\
& -
\frac{ \mathcal J_+^2 - \mathcal J_-^2 }{K_{\bm t}} c_{n+2;y}^\dagger c_{n;y} 
+ \mathrm{H.c.}
-
\sum_n \left( K_{\bm t} + 2 \frac{\mathcal J_+^2 + \mathcal J_-^2 + \Delta_{\rm so}^2 }{K_{\bm t}} \right) c_{n;y}^\dagger c_{n;y}
\, . \label{eqn:Heff_y}
\end{align}
where we have introduced 
$ c^\dag_{n;y} = [ c^\dag_{n;y\uparrow} \; c^\dag_{n;y\downarrow} ]$. 

In the first line of (\ref{eqn:Heff_y}), it emerges the inter-atomic SOI proportional to the SOI strength $\Delta_{\rm so}$, the curvature $\kappa$, the hopping energy $J$, and the reciprocal of the energy of crystal field $K_{\bm t}$, 
\begin{align}
i \frac{2 \mathcal J_- \Delta_{\rm so} }{K_{\bm t}} \tilde{\sigma}_{z}(\phi_n) = i 2 \kappa \alpha \sin(\delta \phi) \frac{ J \Delta_{\rm so} }{K_{\bm t}} \left( {\bm t}(\phi_n) \times {\bm n}(\phi_n) \right) \cdot {\rm \sigma} \, .  
\label{sen:interatomicSOIstrength}
\end{align}
This term is interpreted as the SOI generated by an electric field in the radial direction normal to the helical axis (parallel to ${\bm n}$)~\cite{MatityahuPRB2016}. 
It originates from virtual second-order processes in which a spin transitions to a $\pi$-orbital, simultaneously flipping its direction, both due to SOI, and then hops to a neighboring $\sigma$-orbital due to the misalignment of the local orthogonal coordinates, or, alternatively, the process can occur in the reverse order~\cite{HuertasHernandoPRB2006, VarelaPRB2016}.
In the second line of (\ref{eqn:Heff_y}), the first term represents the second-nearest-neighbor hopping. 
Such long-range hopping processes incorporate interference effects, which are essential for generating the CISS effect in the single-orbital model~\cite{MatityahuPRB2016}.
The second term is the energy gain due to the virtual process.

In the limit of zero torsion, $\tau \to 0$, the bi-normal vector is parallel to the $z$ axis in the original coordinate, 
$\tilde{\sigma}_{z}(\phi_n) = \sigma_z$.  
Then, one can diagonalize the Hamiltonian (\ref{eqn:Heff_y}) by imposing the periodic boundary condition $c_{n+MN} = c_n$ and performing a Fourier transform:  
\begin{align}
{\mathcal H}_{\sigma}^\prime =
\sum_{\ell=0}^{MN-1}  
c_{k_\ell;y}^\dagger {\mathcal H}_{\sigma}^\prime(k_\ell) c_{k_\ell;y} ,  
\;\;
c_{n;y} = \frac{1}{\sqrt{MN}} \sum_{\ell=0}^{MN-1} e^{i k_l n/N} c_{k_\ell;y}\ , \;\; 
k_\ell = \frac{2\pi \ell}{M} \ ,
\end{align}
where the Bloch Hamiltonian is,  
\begin{align}
{\mathcal H}_{\sigma}^\prime(k_\ell) =& \left( -2 J_+ \cos \left( \frac{k_\ell}{N} \right) - K_{\bm t}  
- \frac{4 J_-^2}{K_{\bm t}} \sin^2 \left( \frac{k_\ell}{N} \right)  
- \frac{2 \Delta_{\rm so}^2 }{K_{\bm t}}  
\right) {\sigma}_{0} \notag \\  
&+ \frac{4 J_- \Delta_{\rm so}}{K_{\bm t}} \sin \left( \frac{k_\ell}{N} \right) {\sigma}_{z}  
. \label{eqn:En_sig}  
\end{align}

\section{Results}\label{sec3}

\begin{figure}
\begin{center}
\includegraphics[width=0.9 \columnwidth]{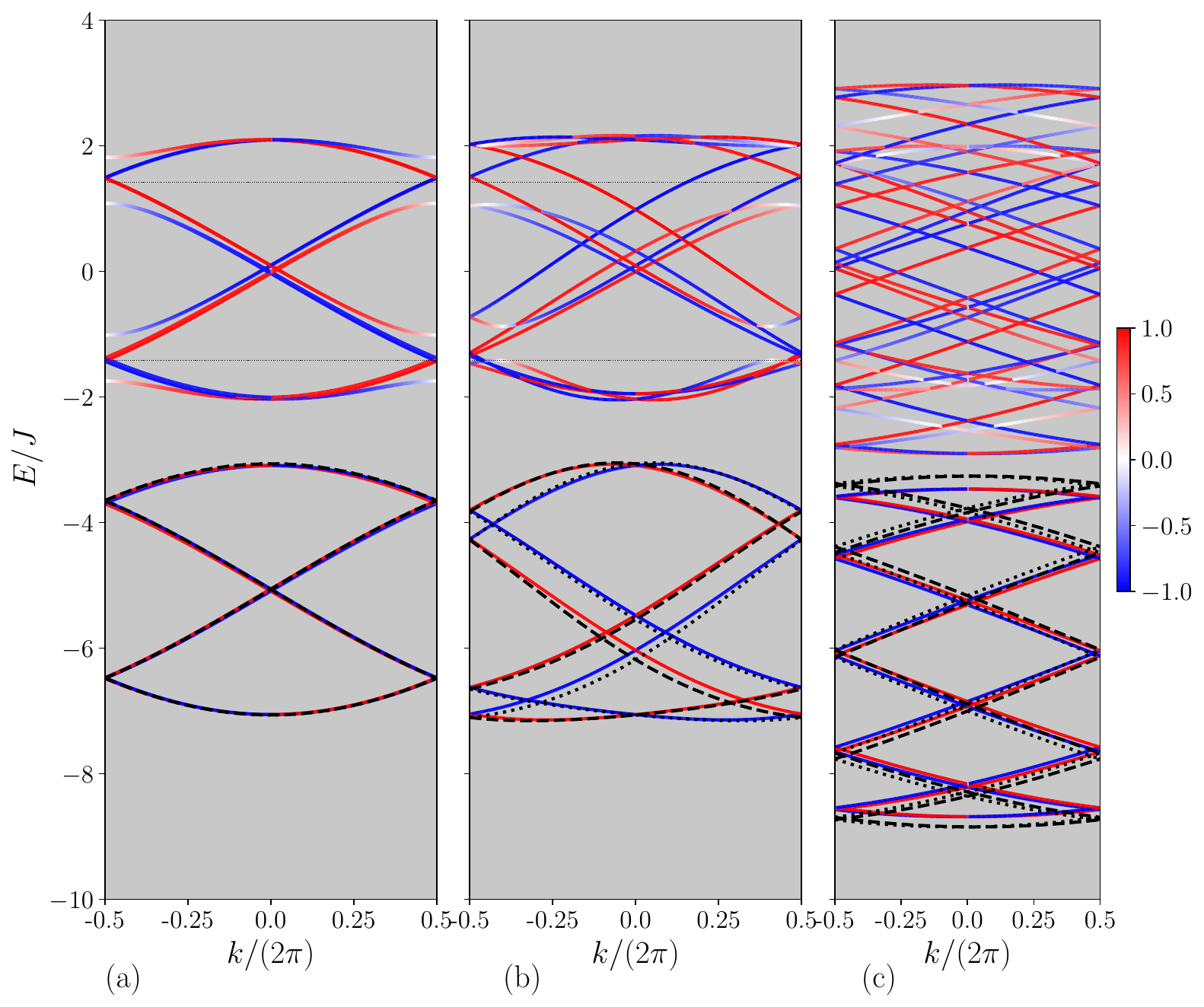}
\caption{
Energy dispersions for  
(a) $\alpha=1$, $\varphi=\Delta \phi$, $\tau \to 0$, $K_{\bm t}=5 J$ and $N=4$,  
(b) $\alpha=\sqrt{2}$, $\varphi=\pi/4$, $\tau \to 0$, $K_{\bm t}=5 J$ and $N=4$,  
and  
(c) $\alpha=\sqrt{2}$, $\varphi=\pi/4$, $\tau=0.48$, $K_{\bm t}=6 J$ and $N=10$.  
The chirality is $p=1$ and the spin-orbit interaction energy is $\Delta_{\rm so}=0.4 J$.  
The dotted and dashed lines indicate the eigenenergies of the $\sigma$-band effective Hamiltonian (\ref{eqn:En_sig}).  
The color scheme indicates the $z$ component of the average spin (red for $\uparrow$ spin and blue for $\downarrow$ spin).
}
\label{fig:band}
\end{center}
\end{figure}

Figures \ref{fig:band} (a-c) show the band structures for various parameters.  
The $\pi$-band (upper band) and the $\sigma$-band (lower band) are separated by the energy of the crystal field, $K_{\bm t}$.  
The color scheme indicates the $z$ component of the average spin (red for $\uparrow$ spin and blue for $\downarrow$ spin).  

The dotted and dashed lines indicate the eigenenergy of the $\sigma$-band effective Hamiltonian (\ref{eqn:En_sig}).  
Panel (a) shows the ideal case: $\alpha=1$, $\varphi=\Delta \phi$, and $\tau \to 0$, where two left-going up spins and two right-going down spins are formed within the energy windows determined by the intra-atomic SOI energy $\Delta_{\rm so}$, the center of which are indicated by the thin horizontal dotted lines at $\mp 2 J \cos (\pi/N)$. 
In this case, $J_-=0$, and thus the inter-atomic SOI in (\ref{eqn:En_sig}) is absent, and the dashed and dotted lines overlap.  
Around the lower thin horizontal dotted line, the degeneracy of each spin pair is slightly lifted because the $\pi$-band is influenced by the $\sigma$-band. 

Panel (b) shows the case away from the ideal condition but still in the limit of zero torsion $\tau \to 0$.  
At the bottom of the $\sigma$-band, we clearly observe Rashba-like spin splitting.  
Overall, the energy dispersion curve is well fitted by (\ref{eqn:En_sig}).  
In the $\pi$-band, the two helical states around the thin horizontal dotted line still remain.  
Note that the size of the energy window is $\Delta_{\rm so}$~\cite{UtsumiPRB2020} and is bigger than the strength of the inter-atomic SOI (\ref{sen:interatomicSOIstrength}).  
In addition, since the $\sigma$-band consists of a single orbital, Rashba-like spin splitting is not sufficient for two-terminal spin filtering~\cite{EntinWohlmanPRB2021}, requiring dissipation to effectively form a multi-terminal system~\cite{MatityahuPRB2016}.

Panel (c) shows the result for parameters chosen to mimic a DNA molecule.  
The number of sites in each turn, $N = 10$, corresponds to the number of base pairs.  
The dimensionless torsion is taken to be $\tau = 0.48$, as estimated for B-form DNA: $R = 1 \mathrm{nm}$ and $\Delta h = 3.4 \mathrm{nm}$~\cite{SasaoJPSJ2019}.  
Due to the finite torsion, the result (\ref{eqn:En_sig}) is no longer applicable, which is the origin of the deviations in panel (c).  
The Rashba-like spin splitting becomes weaker, as it is inversely proportional to the number of sites in the unit cell for small wave numbers, following $\sim 4 J_- \Delta_{\rm so} k_\ell/(K_{\bm t} N)$.

\section{Conclusion}\label{sec4}

By the Schrieffer-Wolff transformation, we derive the inter-atomic SOI starting from the $p$-orbital helical atomic chain.  
For finite curvature, there exists a second-order process where a spin in the $\sigma$-orbital virtually transitions to the $\pi$-band by flipping spin due to the intra-atomic SOI and then hops to the nearest-neighbor $\sigma$-orbital due to the misalignment of $p$-orbitals in adjacent atoms. 
In addition, long-range hopping to the second nearest neighbors emerges. 
Since the virtual process is energetically high due to the crystal field, the effective Hamiltonian well reproduces the spin-split band structure.  
In this way, the model possesses Rashba-like splitting in the $\sigma$-band and helical states in the $\pi$-band simultaneously. 

\backmatter

\bmhead{Acknowledgements}
This work was supported by JSPS KAKENHI Grants No. 20H05666, No. 24K00547, and No. 24K01336. 
This preprint has not undergone peer review (when applicable) or any post-submission improvements or corrections. 
The Version of Record of this article is published in The European Physical Journal Special Topics, and is available online at https://doi.org/10.1140/epjs/s11734-025-02065-1.

\bmhead{Data Availability Statement}
The data that support the findings of this study are available from the corresponding author upon reasonable request.










\begin{appendices}

\section{Spin rotation}
\label{secA}

By introducing the unitary matrix,  
\begin{align}
U^{}_n = e^{i S^{}_x \theta^{}_{p}} e^{i S^{}_z p \, \phi^{}_{n}} \ ,  
\;\;
S_j = \sigma_j / 2 \, ,
\label{Un}
\end{align}
and utilizing the fact that the SOI term conserves the total angular momentum,  
$[ L_i + S_i, {\bm L} \cdot {\bm S} ] = 0$,  
we obtain $[U^{}_{n} O^{}_{n}, {\bm L} \cdot {\bm S}] = 0$,  
which leads to  
\begin{align}
O^{}_n \, {\bm L} \cdot {\bm S} \, O_n^\dagger
= U_n^\dagger U^{}_n O^{}_n \, {\bm L} \cdot {\bm S} \, O_n^\dagger U_n^\dagger U^{}_n
= {\bm L} \cdot U_n^\dagger {\bm S}  U^{}_n \ .
\end{align}
Therefore, (\ref{eqn:tsx}-\ref{eqn:tsz}) are also expressed as  
\begin{align}
\tilde{\sigma}_j(\phi_n) = U_n^\dagger \sigma_j U_n \ , \label{sen:ts}
\end{align}
from which one can check the properties of Pauli matrices,  
$\epsilon_{ijk} \tilde{\sigma}_j \tilde{\sigma}_k = i \tilde{\sigma}_i$  
and  
$\tilde{\sigma}_i^2 = \sigma_0$, 
where $\epsilon_{ijk}$ is the Levi-Civita symbol.

\section{Derivations of (\ref{eqn:Heff_y})}
\label{secB}

We first introduce operators that switch two orbitals: 
\begin{align}
P_x = \ketbra{y}{z} + \ketbra{z}{y} \, , 
\;\;
P_y = \ketbra{z}{x} + \ketbra{x}{z} \, , 
\;\;
P_z = \ketbra{x}{y} + \ketbra{y}{x} \, . 
\end{align}
They satisfy the following relations:  
\begin{align}
[P_x,L_y]_+ =& [P_y,L_x]_+ = - L_z \, , \;\; [P_y,L_z]_+ = [P_z,L_y]_+ = - L_x \, , \notag \\ [P_z,L_x]_+ =& [P_x,L_z]_+ = - L_y \, , 
\\
[L_x, P_y] =& - [L_y, P_x] = i P_z \, , \;\; [L_y, P_z] = - [L_z, P_y] =i P_x \, , \notag \\ [L_z, P_x] =& - [L_x, P_z] = i P_y \, 
\\
[L_x, L_y]_+ =& - P_z \, , \;\; [L_y, L_z]_+ = - P_x \, , \;\; [L_z, L_x]_+ = - P_y \, , 
\\
\left[ P_\alpha, L_\alpha \right]_+ =& 0 \, , 
\;\;
P_\alpha^2 = L_\alpha^2 = I_3 - \ketbra{\alpha}{\alpha} \, .
\end{align}
By exploiting these relations, 
terms in (\ref{eqn:Heff_intermediate}) are calculated as, 
\begin{align}
\left[ \tilde{\bm J}_{\pi-\sigma} , \tilde{\bm J}_{\pi-\sigma}^\dagger \right]_+ 
=&
2 \left( \mathcal{J}_-^2 + \mathcal{J}_+^2 \right) \ketbra{y}{y} + 2 \mathcal{J}_-^2 \ketbra{x}{x} + 2 \mathcal{J}_+^2 \ketbra{z}{z} + 2 \mathcal{J}_+ \mathcal{J}_- P_y \, ,
\\
\left( L_x \tilde{\sigma}_x + L_z \tilde{\sigma}_z \right)^2 
=& 
2 \ketbra{y}{y} + \ketbra{x}{x} + \ketbra{z}{z} - L_y \tilde{\sigma}_y 
\\
\left[ \tilde{\bm J}_{\pi-\sigma} , L_x \tilde{\sigma}_x + L_z \tilde{\sigma}_z  \right]_+ 
=&
-2 i \mathcal{J}_- \ketbra{y}{y} \tilde{\sigma}_z + i \mathcal{J}_- P_{y} \tilde{\sigma}_x + \left( -2 i \mathcal{J}_- \ketbra{x}{x} - \mathcal{J}_+ L_y \right) \tilde{\sigma}_z \, , 
\\
\tilde{\bm J}_{\pi-\sigma} \tilde{\bm J}_{\pi-\sigma}
=&
\left( \mathcal{J}_+^2 - \mathcal{J}_-^2 \right) \ketbra{y}{y} - \mathcal{J}_-^2 \ketbra{x}{x} + \mathcal{J}_+^2 \ketbra{z}{z} - i \mathcal{J}_+ \mathcal{J}_- L_y
\, ,
\end{align}
which lead to (\ref{eqn:Heff_y}).




\end{appendices}


\bibliography{sn-bibliography_fqmt24_utsumi}

\end{document}